\documentclass[12pt]{iopart}
\usepackage{color}
\usepackage{graphicx}
\usepackage{amsmath, amsthm, amssymb}
\usepackage{enumerate}
\usepackage[normalem]{ulem}
\definecolor{Gray}{gray}{0.7}
\usepackage{citesort}
\usepackage{tabularx}
\usepackage{colortbl}
\usepackage[up]{subfigure}

\newcommand{\be}{\begin{equation}}
\newcommand{\ee}{\end{equation}}
\newcommand{\bea}{\begin{eqnarray}}
\newcommand{\eea}{\end{eqnarray}}

\begin{document}

\title[Critical Parameters of the Hard Square Gas]{Estimating the Critical Parameters of the Hard Square Lattice Gas Model} 
\author{Dipanjan Mandal$^1$, Trisha Nath$^2$ and R. Rajesh$^3$}
 \ead{mdipanjan@imsc.res.in$^1$, trisha.nath@theorie.physik.uni-goettingen.de$^2$, rrajesh@imsc.res.in$^3$}
 \address{The Institute of Mathematical Sciences, C.I.T. Campus,
 Taramani, Chennai 600113, India$^{1,3}$}
 \address{Homi Bhabha National Institute, Training School Complex, Anushakti Nagar, Mumbai 400094, India$^{1,3}$}
 \address{Institut f\"{u}r Theoretische Physik, Georg-August-Universit\"{a}t G\"{o}ttingen, 37077
G\"{o}ttingen, Germany$^2$}

\date{\today}

\begin{abstract}
The hard square lattice gas  model on a square lattice is known to undergo a 
continuous phase transition from a low density fluid-like phase to
high density phase with  columnar or smectic order. We estimate the critical activity $z_c$
by calculating, within an approximation scheme, the  interfacial tension between two differently ordered columnar phases,
and then setting it to zero. The approximation scheme allows for the ordered phases to have multiple defects
and the interface between the ordered phases to have overhangs.
We  estimate $z_c=105.35$, which is in 
good agreement with existing Monte Carlo simulation  results of  $z_c \approx 97.5$, and is an improvement
over earlier best estimates of $z_c=54.87$ and $z_c=135.63$. 
\end{abstract}

\maketitle

\section{\label{sec:1}Introduction}

The study of entropy driven transitions in the $2\times2$ hard square lattice gas model, or equivalently
the 2-NN model in which a particle excludes the nearest and next-nearest neighbor from 
being occupied by another particle, has a long history
dating back to the 1950s~\cite{domb,bellerman,hoover,kinzel,amarkaski,francis2,nisbet}. 
The hard square model is known to undergo  a continuous transition from 
a disordered fluid-like phase to an ordered  phase with columnar order as the density $\rho$ or activity $z$ is increased.
The best numerical estimates for the critical behavior, obtained from large scale Monte Carlo simulations, are  critical activity  $z_c\approx 97.5$, critical density $\rho_c \approx 0.932$, and critical
exponents belonging to the Ashkin Teller universality class with critical exponents $\nu \approx 0.92$, $\beta/\nu=1/8$ and
$\gamma/\nu=7/4$~\cite{fernandez,kabir2,feng,zhitomirsky}.  
Unlike the hard hexagon model~\cite{baxter1}, the hard square model is not exactly solvable.
Different analytic and rigorous methods have been used to
estimate the critical parameters over the last few decades~\cite{bellerman,bellerman2,kabir1,lafuente,lafuente2,slotte,trisha2,heitor,temperley2}.
The estimates for  $z_c$ and $\rho_c$ obtained from different methods are summarized in \tref{table:estimates}.
Analytical approaches  like high density expansion~\cite{bellerman,kabir1}, 
Flory-type approximations~\cite{heitor}, density functional theory~\cite{lafuente,lafuente2}, etc.,  result in estimates that 
underestimate the critical activity by more than a factor of 7. Calculations
based on estimating the interfacial tension~\cite{slotte,trisha2} between two ordered phases have been more successful. 
By utilizing the
mapping of the hard square model to the antiferromagnetic Ising model with next nearest neighbor interactions,
a fairly good estimate $z_c=135.63$, that overestimates the critical activity, was obtained, but it is not clear how this 
approach may be extended~\cite{slotte}. In a recent
paper~\cite{trisha2}, we introduced a systematic way of determining the interfacial tension as an expansion in number of defects
in the perfectly ordered phase. While including a single defect improves the estimates for the critical parameters ($z_c=52.49$),
the calculation of the two-defect contribution appears to be too difficult to carry out. We also estimated
the effect of introducing overhangs of height one in the interface for  defect-free phases ($z_c=54.87$). However, it is not clear how 
defects and overhangs may be combined in a single calculation.
In this paper, we determine the interfacial tension using a pairwise approximation, similar
to that used in liquid state theory. This approximation scheme allows us to take into 
account multiple defects as well as overhangs. By determining the
activity at which this interfacial tension vanishes, we  estimate $z_c=105.35$, in reasonable agreement with
numerical results ($z_c\approx 97.5$), and
which is a significant improvement over earlier estimates.
\begin{table}
\caption{\label{table:estimates} 
Estimates of critical activity $z_c$ and critical density $\rho_c$ for columnar-disordered transition of hard square model} 
\begin{indented}
\lineup
\item[]\begin{tabular}{@{}lll}
\br
$z_c$  & $\rho_c$ & Method Used  \\
\mr
   \rowcolor{Gray}
   97.50 &  0.932 & Numerical~\cite{fernandez,kabir2,feng,zhitomirsky} \\
   6.25 &  0.64 & High density expansion (order one)~\cite{bellerman,kabir1}  \\
    11.09 &  0.76  & Flory type mean field~\cite{heitor}\\
     11.09 &  0.76  & Approximate counting~\cite{temperley2}\\
      11.13 &  0.764 & Density Functional theory~\cite{lafuente,lafuente2}\\
     14.86 &  0.754 & High density expansion (order two)~\cite{kabir1}  \\
   17.22 &  0.807 & Rushbrooke Scoins approximation~\cite{bellerman} \\ 
   48.25 & 0.928  & Interfacial tension with no defect~\cite{trisha2}\\
   52.49 & 0.923  & Interfacial tension with one defect~\cite{trisha2}\\
   54.87 & 0.9326 & Interfacial tension with overhang~\cite{trisha2}\\
      135.63&  -     & Interfacial tension in anteferromagnetic Ising model~\cite{slotte}\\
\rowcolor{Gray}
   105.35 &  0.947 & In this paper \\
   \br
  \end{tabular}
  \end{indented}
\end{table}

The hard square model on the square lattice has been studied in different contexts.
It is the prototypical model to study phases with columnar, smectic or layered order
in which translational invariance in broken in some but not all the directions.
Examples of systems showing such ordered phases include liquid crystals~\cite{degennes}, 
adsorbed atoms or molecules on metal surfaces~\cite{adsorbed-on-ni,chlorine,bromine,oxygen,koper},
etc. Columnar phases have also been of recent interest in different hard core lattice gas models.
The hard rectangle gas  shows a nematic-columnar phase transition, in addition to isotropic-nematic
and columnar-sublattice transitions~\cite{joyjit2,joyjit_rectangle_odd}. 
Of these, in the limit of infinite aspect ratio, only the nematic-columnar transition survives  at a finite packing
density~\cite{joyjit3,trisha4}. Generalized models consisting of a mixture of  hard squares and dimers~\cite{kabir2} or 
interacting dimers~\cite{interacting_dimer} also show a columnar phase. The presence of a columnar
phase has also been shown to result in the $k$-NN model, in which the excluded volume of a particle
is made up of its first $k$ next nearest
neighbors, undergoing multiple  entropy driven phase transitions with increasing density~\cite{trisha1,trisha3}.
The study of columnar phases has also been of recent interest in 
quantum spin systems~\cite{quantum_dimer,quantum_dimer2,quantum_dimer3,s_jin,zhitomirsky}. The hard
square system has also found application in modeling adsorption~\cite{chlorine,adsorbed-on-ni}, in combinatorial
problems and tilings~\cite{baxter-comb,squares_torus,packing_square}, and has been the the subject of
recent direct experiments~\cite{brownian_square,vibrating_square}.

The remainder of the paper is organized as follows. In \sref{sec:2}, we define the model precisely and 
outline  the steps in the calculation of the interfacial tension between two ordered columnar phases. The 
calculation involves determining the 
eigenvalue of a transfer matrix $T$, which  is computed in \sref{sec:3}. In \sref{sec:4} the different quantities
determining the largest eigenvalue of $T$ are computed by calculating exactly
the partition function of hard squares on tracks made up of  2 and 4 rows with appropriate boundary conditions. The results
for the interfacial  tension are obtained in \sref{sec:6}. We end with a summary and discussion in \sref{sec:7}.

\section{\label{sec:2}Model and Outline of Calculation}

Consider a square lattice of size $N_x$ $\times$ $N_y$. The sites may be occupied
by particles that are hard squares of size $2\times2$. The squares interact through only
excluded volume interaction i.e. two squares can not overlap but may touch each
other. We associate an activity $z$ to each square.

\begin{figure}
\centering
\includegraphics[width=0.8\columnwidth]{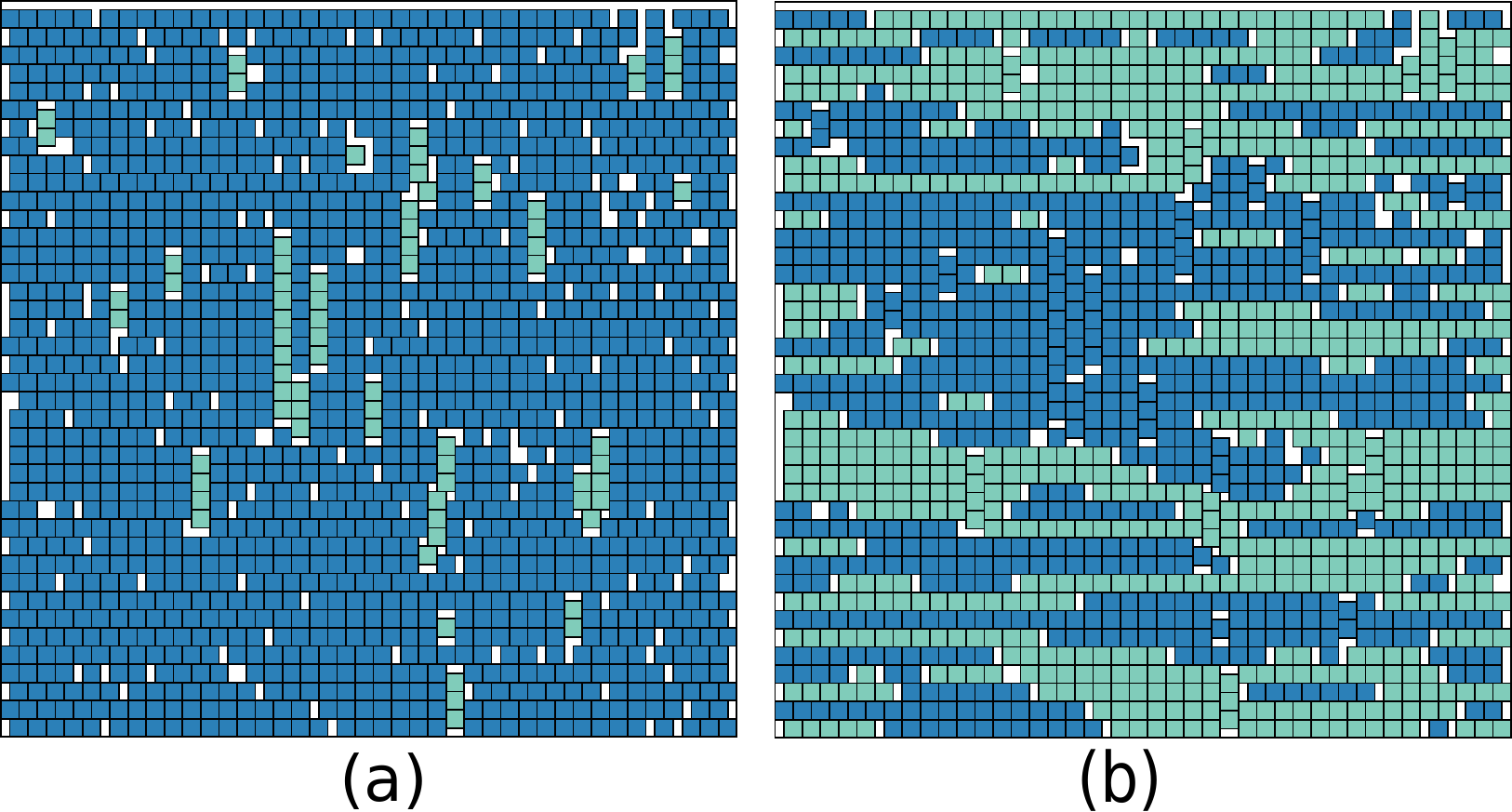}
\caption{Snapshot of a equilibrated configuration of system of hard squares with activity $z=110.0$, corresponding $\rho\approx 0.937$.
These parameters correspond to the system being in an ordered phase. 
A square is colored blue or green depending on whether its head (bottom left point) is in even or odd (a) row and (b) column.
The dominance of one color in (a) implies that the system is a row-ordered phase. The snapshot was generated
using Monte Carlo simulation using the cluster algorithm introduced in \cite{kundu,joyjit1}.}
\label{fig:12}      
\end{figure}

At low activities $z$ or equivalently at low densities $\rho$, the system is in a disordered phase.
For activities larger than critical value $z_c$, the system is in a broken-symmetry phase with 
columnar order, which we define more precisely below. 
Let the  lower left corner of a square be denoted as its head. 
In the columnar phase, the heads preferentially occupy even or odd rows with all columns being equally occupied,
or preferentially occupy even or odd columns with all rows being equally occupied. An example of a row-ordered
phase is shown in \fref{fig:12}. The snapshot of a equilibrated configuration is shown in two different representations.
When the squares are colored according to whether their heads are in even or odd rows [see \fref{fig:12}(a)],
one color is predominantly seen. However, when the same configuration is colored according to whether the
heads of the squares are in even or odd columns [see \fref{fig:12}(b)], then both colors appear in roughly equal proportion.
There are clearly $4$ ordered phases possible.

\begin{figure}
\centering
\includegraphics[width=0.6\columnwidth]{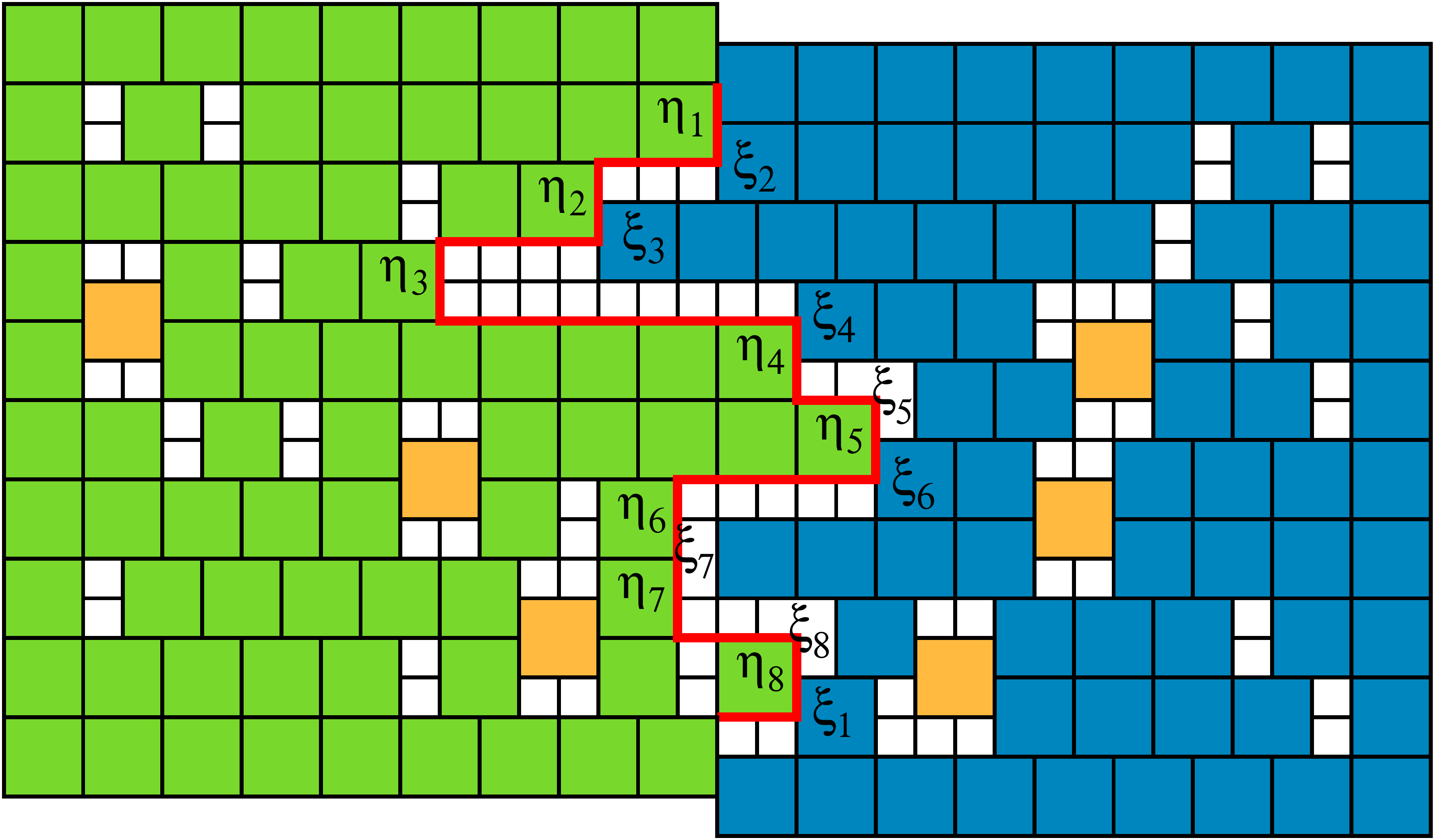}
\caption{Schematic diagram of a configuration in the presence of an interface.
The boundary conditions are such that the left (right) edge of the square is fixed to be occupied
by even (odd) squares. The interface, constituted of the right edges of the right-most squares of the 
left domain is denoted by the red line and labeled by $\eta_i$. $\xi_i$ denotes the left-most position 
possible for a square belonging to the right domain.}
\label{fig:7}      
\end{figure}

The aim of this paper is to estimate the critical activity $z_c$ and critical density $\rho_c$ separating
the disordered phase from the ordered columnar phase. To do so, we determine, within an approximation scheme,
the interfacial tension $\sigma(z)$ between two
differently ordered columnar phase and equate it to zero to obtain the transition point.
Consider boundary conditions where the left edge of the square lattice is fixed to the occupied by
squares with heads in even row and the right edge is fixed to be occupied by squares in
odd row. For large $z$, this choice of boundary condition ensures that 
there is an interface running from top to bottom separating a left phase or domain constituted of squares predominantly 
in even rows from
a right phase or domain constituted of squares predominantly in odd rows. A schematic diagram of the interface is shown in
\fref{fig:7}. We will refer to the two phases as left and
right phases or domains from now on.  Let $Z^{(0)}$ be the partition functions of the system without an interface and 
$Z^{(\mathcal{I})}$ 
be the partition function when an interface $\mathcal{I}$ is present. 
The interfacial tension $\sigma(z)$ is defined as
\be
\label{surf_tension}
e^{-\sigma N_y}=\frac{\sum_{\mathcal{I}}Z^{(\mathcal{I})}}{Z^{(0)}}.
\ee
As  the interactions between the squares are only excluded volume interactions, 
the partition function in the presence of an  interface may be
written as a product of partition function of the left and right phases, i.e.
\be
Z^{(\mathcal{I})}=Z^{(\mathcal{I})}_LZ^{(\mathcal{I})}_R,
\ee
where $Z^{(\mathcal{I})}_L$ and $Z^{(\mathcal{I})}_R$ denote the partition functions of the left  and right phases in the presence of an interface $\mathcal{I}$. It is not possible
to determine $Z^{(\mathcal{I})}_L$, $Z^{(\mathcal{I})}_R$ or $Z^{(0)}$ exactly. In what follows, we calculate these partition functions within  certain
approximations.

First, we assume that the interface between the left and right phases is a directed walk from top to bottom, ie
the interface does not have any upward steps. We define the position of the interface  to be the right boundary of the 
rightmost squares of the left phase. The interface is denoted by $\eta_i$ as shown in \fref{fig:7}. We also
define $\xi_i$ to the left most position that a square in the right phase may occupy on row $i$, as shown in \fref{fig:7}.
Clearly,
\be
\xi_i=\max(\eta_{i-1},\eta_{i}), \quad i=1,2,..,N_y/2. 
\ee
Given an interface, we compute the partition function within an approximation.
The simplest approximation is it to write the partition function as a product of partition functions of tracks of width two,
corresponding to two consecutive rows. This approximation has the drawback that the ordered left and right phases do not have 
any defects, where the squares of wrong type i.e. odd squares in left or even phase and even 
squares in the right or odd phase will be called defects (denoted by yellow in \fref{fig:7}).
The calculation of interfacial tension then reduces to the special case of zero-defects  of \cite{trisha2}. The simplest 
approximation that allows defects to be present is the pairwise approximation, where the partition
function is written as a product of $N_y/2$  partition functions of tracks of width four, made up of four consecutive
rows. We write
\bea
Z^{(\mathcal{I})}_L&=&\frac{\omega_2^{(L)}(\eta_1,\eta_2)\omega_2^{(L)}(\eta_2,\eta_3)~...~\omega_2^{(L)}(\eta_{N_y/2},\eta_1)}
{\mathcal{L}^{(L)}(\eta_1)\mathcal{L}^{(L)}(\eta_2)~...~\mathcal{L}^{(L)}(\eta_{N_y/2})},\\
Z^{(\mathcal{I})}_R&=&\frac{\omega_2^{(R)}(N_x-\xi_1,N_x-\xi_2)~...~\omega_2^{(R)}(N_x-\xi_{N_y/2},
N_x-\xi_1)}
{\mathcal{L}^{(R)}(N_x-\xi_1)~...~\mathcal{L}^{(R)}(N_x-\xi_{N_y/2})},\\
Z^{(0)}&=&\frac{{[\omega_2(N_x,N_x)]}^{N_y/2}}{{[\mathcal{L}(N_x)]}^{N_y/2}},
\label{eq:100}
\eea
where $\omega_2(\ell_1,\ell_2)$ is the partition function of a track of width $4$ where first two rows are of length
$\ell_1$ and third and fourth rows of length $\ell_2$, and $\mathcal{L}(\ell)$ is the partition function of a track of width $2$ 
where both rows have length $\ell$. The superscripts $(L)$ and $(R)$ denote left and right phases. The choice
of the denominator is motivated by the fact that  in the absence of defects,
$\omega_2(\ell_1,\ell_2)= \mathcal{L}(\ell_1) \mathcal{L}(\ell_2)$. In this case,  the overall  partition function should reduce
to a  product over $ \mathcal{L}$'s, and the choice of the denominator ensures this.

The partition functions for the left and right phases are different, and also not the same as the partition function
of the system without an interface, because the presence of the interface imposes  
introduces constraints on the positioning of squares
near the interface. The constraints are as follows. For the left partition function $\omega_2^{(L)}(\ell_1,\ell_2)$, there must
be  even squares (non-defects) present whose right edges are aligned with the position of the interface  in both 
sets of two rows each corresponding to $\ell_1$ and $\ell_2$. This is because the position of the interface has been
defined as the right edge of the rightmost square of the left phase. For the
right partition function $\omega_2^{(L)}(\ell_1,\ell_2)$, the constraint is that there must at least one odd square (non-defect)
between the interface and the left-most defect square. Otherwise, the interface can be redefined to include the defect
square into the left phase. In addition, there is the question of whether defects can be placed between
$\ell_1$ and $\ell_{2}$ for the left and right phases. Placing defects here is equivalent to allowing the interface to have 
overhangs. To prevent overcounting, we will disallow such defects for the left phase, but allow them for the right phase. 
Equivalently, a defect in the left phase may be placed only in the region to the left of $\min(\ell_1, \ell_2)$, and a 
defect in the right phase can be placed to the right of $\min(N_x-\ell_1,N_x-\ell_2)$.

\begin{figure}
\centering
\includegraphics[width=0.4\textwidth]{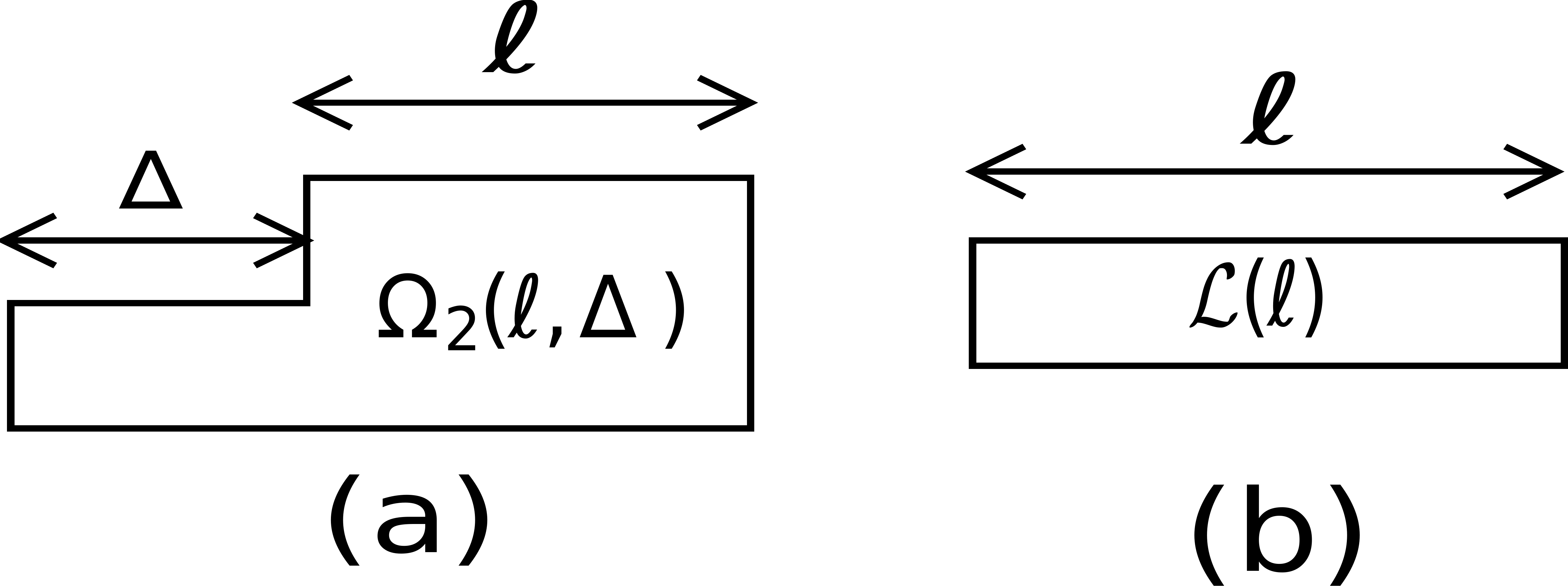}
\caption{Schematic diagram  of a (a) track of width $4$ (four rows) with partition function $\Omega_2(\ell,\Delta)$ 
and (b) track of width $2$ (two rows) with partition function  $\mathcal{L}(\ell)$.}
\label{fig:8}      
\end{figure}

It is convenient to shift to a notation where (see \fref{fig:8})
\be
\omega_2(\ell_1,\ell_2)= \Omega_2\big[\min(\ell_1,\ell_2),|\ell_1-\ell_2|\big]
\ee
Then, the partition function $Z^{(L)}$, $Z^{(R)}$ and $Z^{(0)}$ may be rewritten as
\bea
\label{zp_left}
Z^{(\mathcal{I})}_L
&=&\frac{\prod_{i=1}^{N_y/2}\Omega_2^{(L)}\big[\min(\eta_i,\eta_{i+1}),|\eta_i-\eta_{i+1}|\big]}
{\prod_{i=1}^{N_y/2}\mathcal{L}^{(L)}(\eta_i)},\\
\label{zp_right}
Z^{(\mathcal{I})}_R
&=&\frac{\prod_{i=1}^{N_y/2}\Omega_2^{(R)}\big[N_x-\max(\xi_i,\xi_{i+1}),|\xi_i-\xi_{i+1}|\big]}
{\prod_{i=1}^{N_y/2}\mathcal{L}^{(R)}(N_x-\xi_i)},\\
\label{zp_0}
Z^{(0)}&=&\frac{{[\Omega_2(N_x,0)]}^{N_y/2}}{{[\mathcal{L}(N_x)]}^{N_y/2}},
\eea
For large $\ell$, the partition functions $\Omega_2$ and $\mathcal{L}$ diverge exponentially
with the system size. We define
\bea
\label{12row}
\Omega_2(\ell,\Delta)&=&a_2(\Delta)\lambda_2^{2\ell+\Delta},\\
\label{lrpartitions1}
\Omega_2^{(L)}(\ell,\Delta)&=&a_2^{(L)}(\Delta)\lambda_2^{2\ell+\Delta}, \quad  \ell \gg 1,\\
\label{lrpartitions2}
\Omega_2^{(R)}(\ell,\Delta)&=&a_2^{(R)}(\Delta)\lambda_2^{2\ell+\Delta},
\eea
and
\bea
\label{one_rw}
\mathcal{L}(\ell)&=&a_1 \lambda_1^\ell,\\
\label{lrpartitions3}
\mathcal{L}^{(L)}(\ell)&=&a_1^{(L)} \lambda_1^{\ell},\quad  \ell \gg 1,\\
\label{lrpartitions4}
\mathcal{L}^{(R)}(\ell)&=&a_1^{(R)} \lambda_1^{\ell}.
\eea
Note that we have used the same exponential factor for all $\Omega_2$ (as well as for all $\mathcal{L}$), since 
the free energy is independent of constraints arising from the boundary conditions. It is easy to determine
$a_1^{(L)}$ and $a_1^{(R)}$ in terms of $a_1$. In the left domain, for a track of width 2, the constraint is that the rightmost
square must touch the interface. This means that $\mathcal{L}^{(L)}(\ell)=z\mathcal{L}(\ell-2)\approx za_1\lambda_1^{\ell-2}
$. In the right domain, defects cannot be present in a track of width 2, and hence there are no constraints,
implying that $\mathcal{L}^{(R)}(\ell)=\mathcal{L}(\ell) \approx a_1\lambda_1^{\ell}$. Therefore,
\bea
\label{lrpartitions}
a_1^{(L)}&=&\frac{z a_1}{\lambda_1^2},\\
a_1^{(R)}&=&a_1.
\eea

Using the asymptotic forms for the partition functions, the 
partition functions of the left [see \eref{zp_left}] and right [see \eref{zp_right}] phases may 
be rewritten as
\bea
Z^{(\mathcal{I})}_L&=&\frac{\prod_{i=1}^{N_y/2}a_2^{(L)}\big(|\eta_i-\eta_{i+1}|\big)
\lambda_2^{2\min(\eta_i,\eta_{i+1})+|\eta_i-\eta_{i+1}|}}
{\prod_{i=1}^{N_y/2}za_1\lambda_1^{\eta_i-2}},\\
Z^{(\mathcal{I})}_R&=&\frac{\prod_{i=1}^{N_y/2}a_2^{(R)}\big(|\xi_i-\xi_{i+1}|\big)
\lambda_2^{2N_x-2\max(\xi_i,\xi_{i+1})+|\xi_i-\xi_{i+1}|}}
{\prod_{i=1}^{N_y/2}a_1\lambda_1^{N_x-\xi_i}}.
\eea
Using the relations $2\min(m,n)=m+n-|m-n|$ and $2\max(m,n)=m+n+|m-n|$,  taking
product of $Z^{(L)}$ and $Z^{(R)}$ and simplifying, we obtain
\be
\label{approxzi}
Z^{(\mathcal{I})}=\frac{\lambda_2^{N_xN_y}\prod_{i=1}^{N_y/2}a_2^{(L)}\big(|\eta_i-\eta_{i+1}|\big)
a_2^{(R)}\big(|\xi_i-\xi_{i+1}|\big)\lambda_2^{-|\eta_i-\eta_{i+1}|}}{{\bigg(\frac{z a_1^2}{\lambda_1^2}
\bigg)}^{N_y/2}\lambda_1^{N_xN_y/2}\prod_{i=1}^{N_y/2}\lambda_1^{-\frac{1}{2}|\eta_i-\eta_{i+1}|}}.
\ee
Likewise, the partition function of the system without an interface [see \eref{zp_0}] may be written for 
large $N_x$ as
\be
\label{z0}
Z^{(0)}={\bigg[\frac{a_2(0)\lambda_2^{2N_x}}{a_1\lambda_1^{N_x}}\bigg]}^{N_y/2}.
\ee
Knowing the partition functions \eref{approxzi} and \eref{z0}, the interfacial tension in \eref{surf_tension} may be expressed
in terms of $a$'s, $\lambda_1$ and $\lambda_2$ as
\be
\label{approxsigma}
e^{-\sigma N_y}={\bigg[\frac{\lambda_1^2}{z a_1 a_2(0)}\bigg]}^{N_y/2}\sum_{\mathcal{I}}\prod_{i=1}^{N_y/2}
\frac{a_2^{(L)}\big(|\eta_i-\eta_{i+1}|\big)
a_2^{(R)}\big(|\xi_i-\xi_{i+1}|\big)\lambda_2^{-|\eta_i-\eta_{i+1}|}}{\lambda_1^{-\frac{1}{2}|\eta_i-\eta_{i+1}|}}.
\ee
We note that all arguments are in terms of differences between consecutive $\eta_i$'s or $\xi_i$'s. It is therefore
convenient to introduce new variables
\be
{\widetilde{\eta}}_i=\eta_i-\eta_{i-1}.
\ee
In terms of these new variables, it is straightforward to derive
\be
\xi_{i+1}-\xi_i={\widetilde{\eta}}_{i+1}\theta({\widetilde{\eta}}_{i+1})+{\widetilde{\eta}}_{i}
(1-\theta({\widetilde{\eta}}_{i})),
\ee
where $\theta(x)$ is the Heaviside step function defined as
$\theta(x)=1$ for $x \geq 0$ and $\theta(x)=0$ for $x<0$.
In terms of these new variables ${\widetilde{\eta}}_i$,
the interfacial tension \eref{approxsigma} may be rewritten as
\bea
\label{approxsigma2}
e^{-\sigma N_y}={\bigg[\frac{\lambda_1^2}{z a_1 a_2(0)}\bigg]}^{N_y/2}\sum_{[{\widetilde{\eta}}_{i}]}\prod_{i=1}^{N_y/2}
&{\bigg(\frac{\sqrt{\lambda_1}}{\lambda_2}\bigg)}^{|{\widetilde{\eta}}_{i}|}
a_2^{(L)}\big(|{\widetilde{\eta}}_{i}|\big)\times\nonumber\\
&a_2^{(R)}\big(|{\widetilde{\eta}}_{i+1}\theta({\widetilde{\eta}}_{i+1})+{\widetilde{\eta}}_{i}(1-\theta({\widetilde{\eta}}_{i}))|\big),
\eea
where the sum over ${\widetilde{\eta}}_{i}$ varies from $-\infty$ to $+\infty$. 

The summation over ${\widetilde{\eta}}_{i}$  is not straightforward to do as they are not independent due to terms 
coupling ${\widetilde{\eta}}_{i}$ and ${\widetilde{\eta}}_{i+1}$. To do the sum, we define an
infinite dimensional transfer matrix $T$  with coefficients
\be
\label{transfer_matrix}
T_{{\widetilde{\eta}}_{i},{\widetilde{\eta}}_{i+1}}={\bigg(\frac{\sqrt{\lambda_1}}{\lambda_2}\bigg)}^{|{\widetilde{\eta}}_{i}|}
a_2^{(L)}\big(|{\widetilde{\eta}}_{i}|\big)
a_2^{(R)}\big(|{\widetilde{\eta}}_{i+1}\theta({\widetilde{\eta}}_{i+1})+{\widetilde{\eta}}_{i}(1-\theta({\widetilde{\eta}}_{i}))|\big).
\ee
Let $\Lambda_2$ be the largest eigenvalue of the transfer matrix $T$. For large $N_y$, we may then write \eref{approxsigma2} as
\be
\label{sigma1}
e^{-\sigma N_y}={\bigg[\frac{\lambda_1^2}{z a_1 a_2(0)}\bigg]}^{N_y/2}\sum_{[{\widetilde{\eta}}_{i}]}\prod_{i=1}^{N_y/2}
T_{{\widetilde{\eta}}_{i},{\widetilde{\eta}}_{i+1}}
={\bigg[\frac{\lambda_1^2\Lambda_2}{z a_1 a_2(0)}\bigg]}^{N_y/2}.
\ee
At the transition point, $\sigma$ vanishes, and the critical activity $z_c$ therefore satisfies the relation
\be
\label{critical_condition}
\frac{\lambda_1^2\Lambda_2}{z_c a_1 a_2(0)}=1,
\ee
where $\Lambda_2$ depends on $a_2^{(R)}$ and $a_2^{(L)}$. These unknown parameters are calculated exactly
in \sref{sec:3} and \sref{sec:4}.

\section{\label{sec:3} Calculation of Eigenvalue of $T$}

In this section, we determine the largest eigenvalue of the transfer matrix $T$ with components as defined
in \eref{transfer_matrix}. Let the largest eigenvalue of $T$ be denoted by $\Lambda_2$ corresponding to
an eigenvector $\Psi$ with components $\psi_i$, $i=-\infty, \ldots, \infty$.
In component form, the eigenvalue equation is
\be
\label{eigenvalue_eqn}
\sum_{j=-\infty}^{\infty}T_{i,j}\psi_j=\Lambda_2\psi_i, ~i=-\infty, \ldots, \infty.
\ee
Substituting for $T$ from \eref{transfer_matrix}, we obtain
\be
\label{jgeq0}
\bigg(\frac{\sqrt{\lambda_1}}{\lambda_2}\bigg)^{|i|}a_2^{(L)}(|i|)\bigg[a_2^{(R)}(0)\sum_{j=-\infty}^0\psi_j+\sum_{j=1}^{\infty}a_2^{(R)}(|j|)\psi_j\bigg]\\ =\Lambda_2 \psi_i,~ i \geq 0,
\ee
\be
\label{jl0}
\bigg(\frac{\sqrt{\lambda_1}}{\lambda_2}\bigg)^{|i|}a_2^{(L)}(|i|)\bigg[a_2^{(R)}(|i|)\sum_{j=-\infty}^{0}\psi_j+\sum_{j=1}^{\infty}a_2^{(R)}(|j+i|)\psi_j\bigg]\\=\Lambda_2 \psi_i,~i < 0.
\ee

First consider the case for $i\geq0$. \Eref{jgeq0} may be re-written as
\be
\label{psigeq0}
\bigg(\frac{\sqrt{\lambda_1}}{\lambda_2}\bigg)^{|i|}a_2^{(L)}(|i|)\bigg[a_2^{(R)}(0)\beta+\sum_{j=1}^{\infty}a_2^{(R)}(|j|)\widetilde{\psi}_j\bigg]=\Lambda_2 \widetilde{\psi}_i, ~i \geq 0,
\ee
where 
\be
\widetilde{\psi}_i = \frac{\psi_i}{\psi_0}; ~~\beta = \sum_{i=-\infty}^{0}\widetilde{\psi}_i.
\label{beta}
\ee
Since $\widetilde{\psi}_0=1$, from \eref{psigeq0} with $i=0$, we immediately obtain the eigenvalue $\Lambda_2$ to be
\be
\label{Lambda}
\Lambda_2=a_2^{(L)}(0) \left[ a_2^{(R)}(0)\beta+\sum_{j=1}^{\infty} a_2^{(R)}(|j|)\widetilde{\psi}_j\right].
\ee
with components of the eigenvector being
\be
\label{solution1}
\widetilde{\psi}_i={\bigg(\frac{\sqrt{\lambda_1}}{\lambda_2}\bigg)}^{|i|}\frac{a_2^{(L)}(|i|)}{a^{(L)}(0)},~i \geq 0.
\ee

Now, consider the case $i<0$. In terms of $\widetilde{\psi}_i$, \eref{jl0} may be written as
\be
{\bigg(\frac{\sqrt{\lambda_1}}{\lambda_2}\bigg)}^{|i|}a_2^{(L)}(|i|)\bigg[a_2^{(R)}(|i|)\beta+\sum_{j=1}^{\infty}a_2^{(R)}(|j+i|)\widetilde{\psi}_j\bigg]=\Lambda_2 \widetilde{\psi}_i,~i < 0.
\ee
Substituting $\widetilde{\psi}_j$ for $j\geq 0$, from \eref{solution1}, we obtain
\be
\label{solution2}
{\bigg(\frac{\sqrt{\lambda_1}}{\lambda_2}\bigg)}^{|i|}a_2^{(L)}(|i|) F(i) =
\Lambda_2\widetilde{\psi}_i,~i< 0,
\ee
where, the function $F(i)$ is defined as
\be
F(i)=a_2^{(R)}(|i|)\beta+\sum_{j=1}^{\infty}{\bigg(\frac{\sqrt{\lambda_1}}{\lambda_2}\bigg)}^{|j|}\frac{a_2^{(R)}(|j+i|)a_2^{(L)}(|j|)}{a_2^{(L)}(0)}.
\ee
The solution to \eref{solution2} is clearly
\be
\Lambda_2=a_2^{(L)}(0)F(0),
\ee
which is consistent with \eref{Lambda}, and
\be
\label{solution3}
\widetilde{\psi}_i={\bigg(\frac{\sqrt{\lambda_1}}{\lambda_2}\bigg)}^{|i|}\frac{a_2^{(L)}(|i|)F(i)}{a_2^{(L)}(0)F(0)},~i< 0.
\ee
\Eref{Lambda}, \eref{solution1}, and \eref{solution3} determine $\Lambda_2$ and the components of the eigenvector.
To solve for $\Lambda_2$ in terms of $a_2^{(L)}(\Delta)$ and $a_2^{(R)}(\Delta)$, it is convenient to define three 
quantities 
\bea
\label{k1}
k_1&=&\sum_{i=1}^{\infty}{\bigg(\frac{\sqrt{\lambda_1}}{\lambda_2}\bigg)}^{|i|}a_2^{(L)}(|i|)a_2^{(R)}(|i|),\\
\label{k2}
k_2&=&\sum_{i=-\infty}^{0}{\bigg(\frac{\sqrt{\lambda_1}}{\lambda_2}\bigg)}^{|i|}a_2^{(L)}(|i|)a_2^{(R)}(|i|),\\
\label{k3}
k_3&=&\sum_{i=-\infty}^{0}\sum_{j=1}^{\infty}{\bigg(\frac{\sqrt{\lambda_1}}{\lambda_2}\bigg)}^{|i|+|j|}\frac{a_2^{(L)}(|i|)a_2^{(R)}(|i+j|)a_2^{(L)}(|j|)}{a_2^{(L)}(0)}.
\eea
Solving for $\beta$ in \eref{beta} and \eref{Lambda} by substituting for  $\widetilde{\psi}_i$ from \eref{solution3} and
\eref{solution1} respectively, we obtain
\bea
\label{beta1}
\beta&=&\frac{k_3}{\Lambda_2-k_2},\\
\beta&=&\frac{\Lambda_2-k_1}{a_2^{(L)}(0)a_2^{(R)}(0)}.
\label{beta2}
\eea
Equating \eref{beta1} and \eref{beta2} to eliminate $\beta$, we find that $\Lambda_2$ satisfies the quadratic equation
\be
\label{eq_of_lambda}
\Lambda_2^2-(k_1+k_2)\Lambda_2 +k_1k_2-k_3a_2^{(L)}(0)a_2^{(R)}(0)=0,
\ee
whose largest root is
\be
\label{largest_lambda}
\Lambda_2=\frac{k_1+k_2+\sqrt{{(k_1-k_2)}^2+4a_2^{(L)}(0)a_2^{(R)}(0)k_3}}{2}.
\ee
The largest eigenvalue may be further simplified using 
\bea
\label{ktilda1}
&k_2-k_1&=a_2^{(L)}(0)a_2^{(R)}(0),\\
\label{ktilda2}
&\widetilde{k}&=k_2+k_1=\sum_{i=-\infty}^{\infty}{\bigg(\frac{\sqrt{\lambda_1}}{\lambda_2}\bigg)}^{|i|}a_2^{(L)}(|i|)a_2^{(R)}(|i|).
\eea
After simplification we get the largest eigenvalue
\be
\label{largest_lambda_s}
\Lambda_2=\frac{\widetilde{k}+\sqrt{{\big[a_2^{(L)}(0)a_2^{(R)}(0)\big]}^2+4a_2^{(L)}(0)a_2^{(R)}(0)k_3}}{2},
\ee
with $k_3$ as in \eref{k3} and $\widetilde{k}$ as in \eref{ktilda2}.

\section{\label{sec:4}Calculation of Partition Functions of Tracks}

\subsection{Partition function of track of width $2$}

\begin{figure}
\centering
\includegraphics[width=0.6\columnwidth]{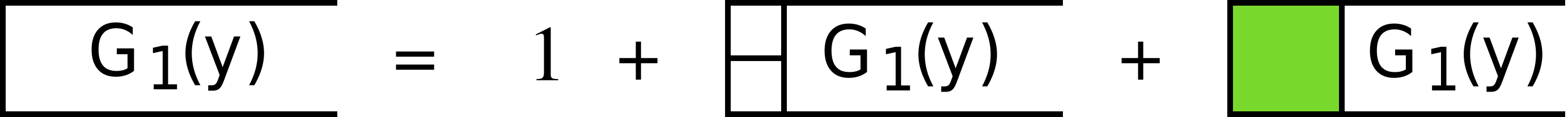}
\caption{Diagrammatic representation of the recursion relation obeyed by the generating function
$G_1(y)$ defined for a track of width 2 [see \eref{g1y} for definition]. The first column of the track
may be occupied by two vacancies (open 1$\times$1 square) or a square (filled 2$\times$2 square).}
\label{fig:15}      
\end{figure}

In this section, we determine the asymptotic behavior of the partition function $\mathcal{L}(\ell)$ of a
track of width $2$ and length $\ell$ [the shape  of the track is shown in \fref{fig:8}(b)]. We define the generating function
\be
\label{g1y}
 G_1(y)=\sum_{\ell=0}^\infty \mathcal{L}(\ell)y^{\ell},
\ee
where the power of $\sqrt{y}$ is the number of sites present in the system.
The recursion relation obeyed by $G_1(y)$ is shown diagrammatically in \fref{fig:15} and 
can be written as
\be
G_1(y)=1+yG_1(y)+zy^2G_1(y),
\ee
which may be solved to give
\be
\label{G1y}
G_1(y)=\frac{1}{1-y-zy^2}.
\ee
Let $y_1$ be the smallest root of the denominator $1-y-zy^2$ of \eref{G1y}, i.e.
\be
y_1=\frac{\sqrt{1+4z}-1}{2z}.
\ee
By finding the coefficient of $y^\ell$ for large $\ell$, it is straightforward to obtain
\be
\label{mathcalL}
\mathcal{L}(\ell)=a_1\lambda_1^{\ell}[1+O(\exp(-c\ell))],~c>0,~\ell \gg 1,
\ee
where
\be
\label{lmda1a1}
\lambda_1=\frac{1}{y_1},~~a_1=\frac{1}{2-y_1}.
\ee

\subsection{Partition functions for tracks of width $4$}

In this section we determine the partition functions of tracks of width $4$ without any constraints. The shape
of a generic track of width $4$ is characterized by parameters $\ell$ and $\Delta$, and is shown in \fref{fig:8} (a).
Calculating these partition functions will allow us to  determine $a_2(\Delta)$ as defined in \eref{12row}. 
\begin{figure}
\centering
\includegraphics[width=0.6\columnwidth]{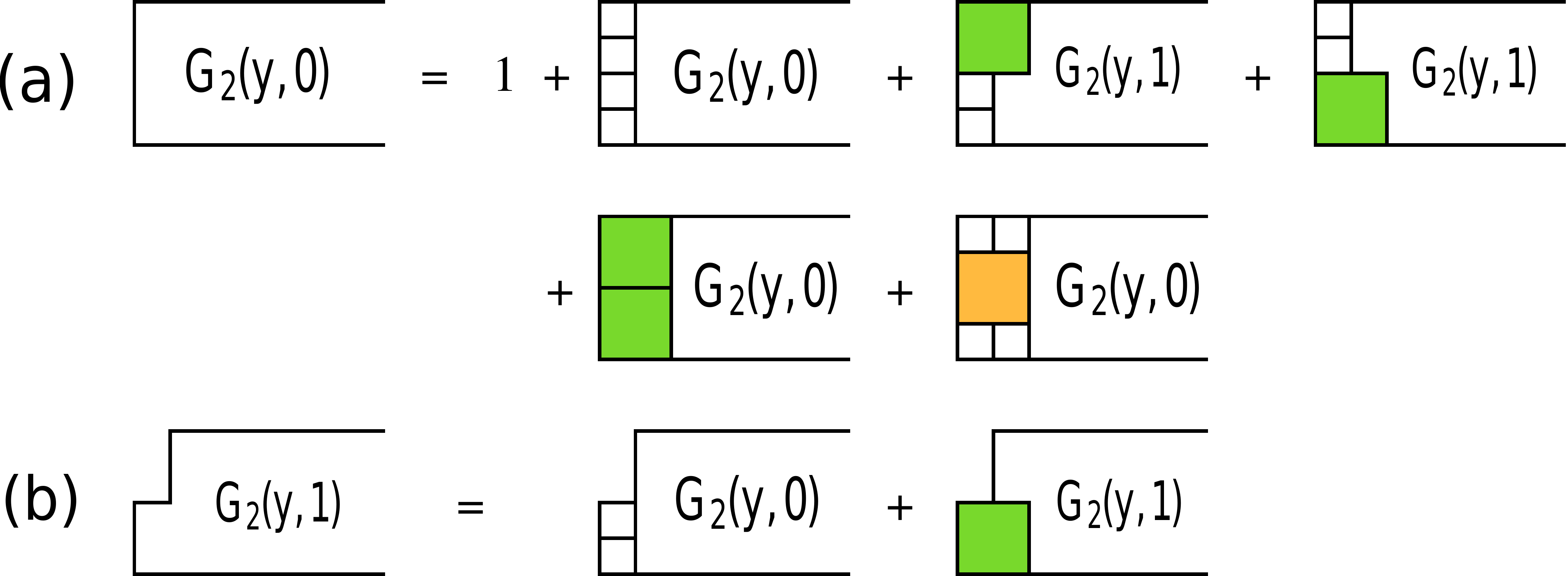}
\caption{Diagrammatic representation of the recursion relation obeyed by the generating functions
(a) $G_2(y,0)$ and (b) $G_2(y,1)$ for a track of width 4 [see \eref{g2delta} for definition]. Right
hand side enumerates the different ways the first column of the track may be occupied by vacancies (open 1$\times$1 square),
square (filled 2$\times$2 green square) and defect (filled 2$\times$2 yellow square).}
\label{fig:1}      
\end{figure}

Consider the following generating function.
\be
\label{g2delta}
 G_2(y,\Delta)=\sum_{\ell=0}^\infty \Omega_2(\ell,\Delta)y^{2\ell+\Delta},
\ee
where the power of $\sqrt{y}$ is the number of sites in the system. 
$G_2(y,0)$ and $G_2(y,1)$ obey simple recursion relations which are shown 
diagrammatically
in \fref{fig:1}. In equation form, they are  
\bea
\label{g20_00}
 G_2(y,0)=&1+y^2G_2(y,0)+2zy^3G_2(y,1)+(z^2y^4 +z_Dy^4)G_2(y,0),\\
 \label{g20_11}
 G_2(y,1)=&yG_2(y,0)+zy^2G_2(y,1),
\eea
where $z_D$ is the activity associated with each defect square.
These relations are easily solved to give
\bea
\label{sg20}
 G_2(y,0)&=&\frac{1-zy^2}{f(y^2)},\\
 \label{sg21}
 G_2(y,1)&=&\frac{y}{f(y^2)},
\eea where
\be
f(y)=z(z^2+z_D)y^3-(z^2+z+z_D)y^2-(1+z)y+1.\nonumber
\ee
Let $y_2$ be the smallest root of $f(y)=0$. For very large $\ell$, we may write
$\Omega_2(\ell,\Delta)$ as
\be
\label{approx_omega}
\Omega_2(\ell,\Delta)=a_2(\Delta)\lambda_2^{2\ell+\Delta}[1+O(\exp(-c\ell))],~\ell \gg 1, ~c>0,
\ee
where
\be
\lambda_2=\frac{1}{\sqrt{y_2}}.
\ee
Calculating coefficient of $y^{2\ell+\Delta}$, the prefactor $a_2(\Delta)$ for
$\Delta=0,1$ is obtained to be
\bea
\label{aa20}
a_2(0)&=&\frac{-(1-zy_2)}{y_2f^\prime(y_2)},\\
\label{aa21}
a_2(1)&=&\frac{-1}{\sqrt{y_2}f^\prime(y_2)}.
\eea

\begin{figure}
\centering
\includegraphics[width=0.5\columnwidth]{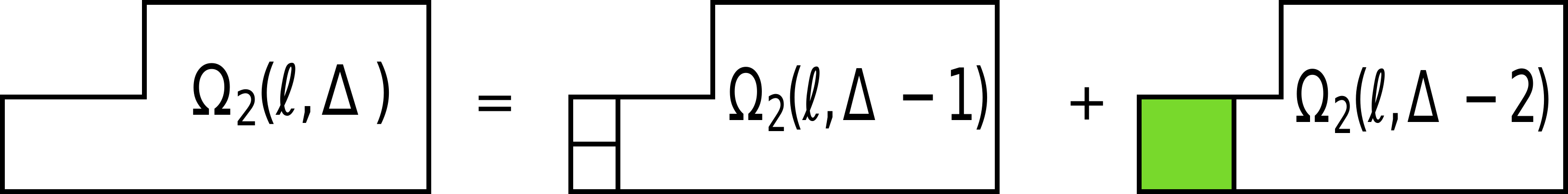}
\caption{Diagrammatic representation of the recursion relation obeyed by the partition
function $\Omega_2(\ell,\Delta)$ with $\Delta\geq2$, for a track of width 4. The first column of the
track may be occupied by two vacancies (open 1$\times$1 square) or a square (filled 2$\times$2 square).}
\label{fig:3}      
\end{figure}

We now consider $\Delta \geq 2$. The recursion relation obeyed by $\Omega_2(\ell,\Delta)$ for $\Delta \geq 2$ is shown diagrammatically
in \fref{fig:3}, and may be written mathematically as
\be
\label{o2ld}
\Omega_2(\ell,\Delta)=\Omega_2(\ell,\Delta-1)+z\Omega_2(\ell,\Delta-2),~~\Delta=2, 3,...
\ee
We define the generating function
\be
\label{flx}
F(\ell,x)=\sum_{\Delta=0}^\infty \Omega_2(\ell,\Delta)x^{\Delta}.
\ee
Multiplying \eref{o2ld} by $x^\Delta$ and summing from $2$ to $\infty$, we obtain a linear equation obeyed by $F(\ell,x)$ which is 
easily solved to give
\be
F(\ell,x)=\frac{\Omega_2(\ell,0)+x\big[\Omega_2(\ell,1)-\Omega_2(\ell,0)\big]}{1-x-zx^2},
\ee
where $\Omega_2(\ell,0)$ and $\Omega_2(\ell,1)$ have already been determined [see \eref{sg20}, \eref{sg21}].
$F(\ell,x)$ has two simple poles at
\be
x_\pm=\frac{-1\pm\sqrt{1+4z}}{2z}.
\ee
Expanding the denominator about its two roots $x_\pm$, we determine $\Omega_2(\ell,\Delta)$ by calculating the coefficient of 
$x^\Delta$. We obtain
\be
\label{a2dlta}
a_2(\Delta)=A_+(x_+\lambda_2)^{-\Delta}+A_-(x_-\lambda_2)^{-\Delta},~\Delta= 0,1,2...,
\ee
where
\be
A_\pm=\frac{\pm\big[\lambda_2 a_2(1)-(z x_\mp +1)a_2(0)\big]}{\sqrt{1+4z}}.
\ee

\subsection{Calculation of $a_2^{(L)}(\Delta)$}

In this section, we calculate the pre-factor $a_2^{(L)}(\Delta)$ that characterizes the asymptotic behavior of the partition
function of track of width 4 [see \eref{lrpartitions1}] for the left phase. The left phase has the constraint that the right edge 
of the rightmost square must touch the interface [see discussion in the paragraph following  \eref{eq:100}]. Thus
\be
\label{omega2_left}
\Omega_2^{(L)}(\ell,\Delta)=z^2\Omega_2(\ell-2,\Delta),
\ee
where the factor $z^2$ accounts for the two squares adjacent to interface. Once these two squares are placed the 
occupation of the rest of the track has no constraints and hence enumerated by $\Omega_2(\ell-2,\Delta)$.
Using \eref{omega2_left}, \eref{lrpartitions1} and \eref{approx_omega}, for very large $\ell$ we obtain
\be
\label{a2_left}
a_2^{(L)}(\Delta)=\frac{z^2}{\lambda_2^4} a_2(\Delta),
\ee
where $a_2(\Delta)$ is given in \eref{a2dlta}.

\subsection{Calculation for $a_2^{(R)}(\Delta)$ \label{sec:secar}}

In this section, we calculate $a_2^{(R)}(\Delta)$ for $\Delta \geq 0$,
as defined in \eref{lrpartitions2}. Consider the track labeled by $(\xi_i,\xi_{i+1})$ [see \fref{fig:7}].
The constraint on the right phase is that 
a defect is allowed to be present only to to the right of $\min(\xi_i,\xi_{i+1})$ and there must be at least one
non-defect square present to its left [see discussion in the paragraph following  \eref{eq:100}]..

\begin{figure}
\centering
\includegraphics[width=0.5\columnwidth]{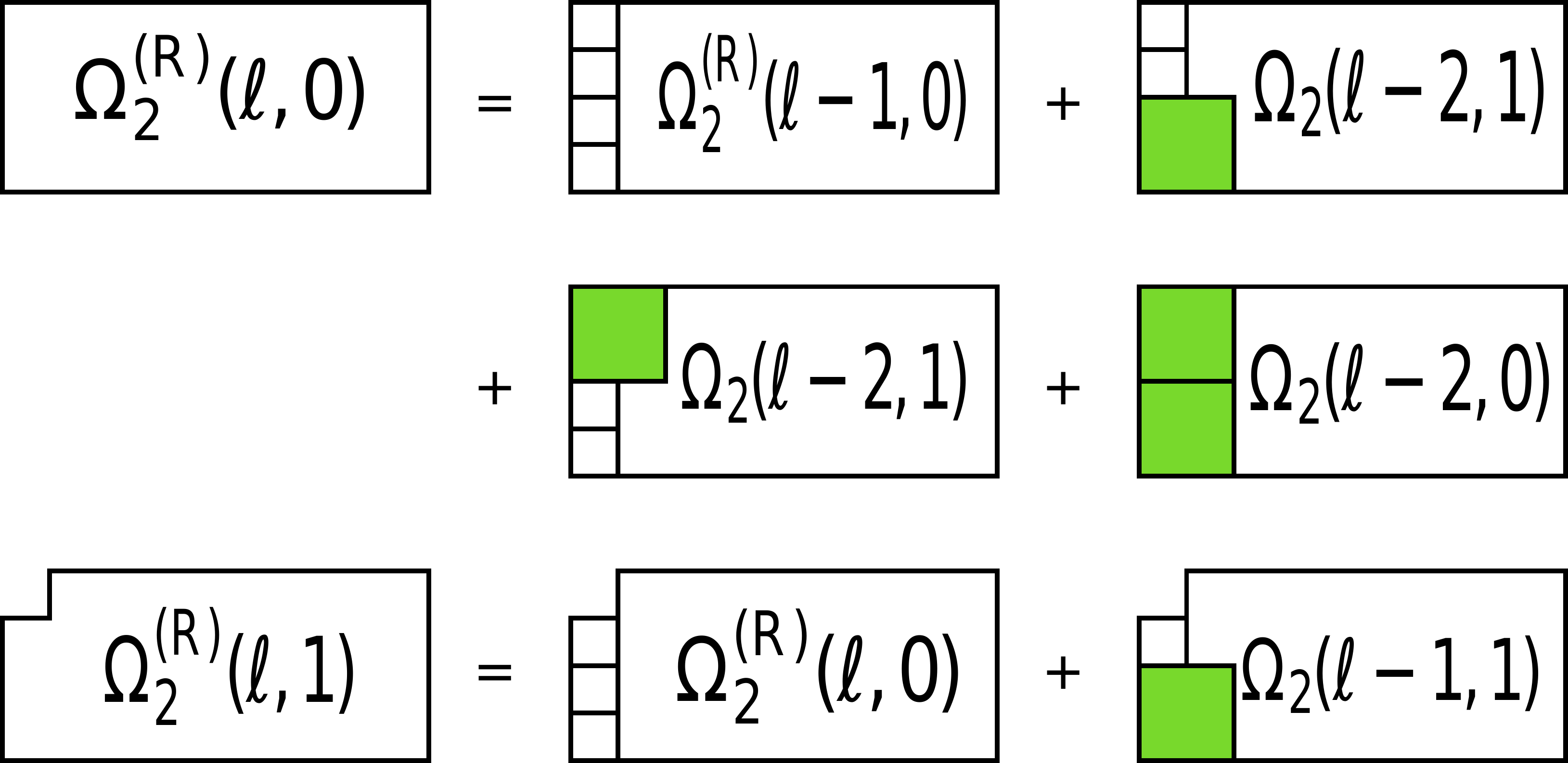}
\caption{Diagrammatic representation of the recursion relation obeyed by the partition
functions $\Omega_2^{(R)}(\ell,0)$ and $\Omega_2^{(R)}(\ell,1)$ for the track of width 4. 
Right hand side enumerates the different ways the first 
column of the track may be occupied by vacancies (open 1$\times$1 square) or squares (filled 2$\times$2 square).}
\label{fig:4}      
\end{figure}

First consider $\Delta=0,1$.
The recursion relation obeyed by the partition functions $\Omega_2^{(R)}(\ell,0)$ and
$\Omega_2^{(R)}(\ell,1)$ for right phase are shown diagrammatically in \fref{fig:4} and may be
written as
\bea
\Omega_2^{(R)}(\ell,0)&=&\Omega_2^{(R)}(\ell-1,0)+2z\Omega_2(\ell-2,1)+z^2\Omega_2(\ell-2,0),\\
\Omega_2^{(R)}(\ell,1)&=&\Omega_2^{(R)}(\ell,0)+z\Omega_2(\ell-1,1).
\eea
Using the asymptotic expressions for the partition functions as given in \eref{12row} and \eref{lrpartitions2}, 
we obtain two linear equations 
for $a_2^{(R)}(0)$ and $a_2^{(R)}(1)$, which are easily solved to give
\bea
\label{a2RL}
a_2^{(R)}(0)&=&\frac{z\big[2a_2(1)\lambda_2+za_2(0)\big]}{\lambda_2^2(\lambda_2^2-1)},\\
a_2^{(R)}(1)&=&\frac{\lambda_2a_2^{(R)}(0)+za_2(1)}{\lambda_2^2}.
\eea

Now consider $\Delta \geq 2$.
The recursion relation obeyed by $\Omega_2^{(R)}(\ell,\Delta)$
for $\Delta \geq 2$ may be written as
\be
\label{omega2Relldelta0}
\Omega_2^{(R)}(\ell,\Delta)=\Omega_2^{(R)}(\ell,\Delta-1)+z\widetilde{\Omega}_2(\ell,\Delta-2)
\ee
where $\widetilde{\Omega}_2(\ell,\Delta)$ is the partition function for a generalization of the shape 
for $\Omega_2^{(R)}(\ell,1)$ in the left hand side of \fref{fig:4}. The lack of the subscript $(R)$ means
that there are no constraints. The first term in the right hand side of \eref{omega2Relldelta0} 
corresponds to placing vacancies in first column, and the second term to a non-defect square being placed.
$\Omega_2^{(R)}(\ell,\Delta-1)$ in the right hand side of \eref{omega2Relldelta0} may be iterated 
further to yield
\be
\label{omega2Relldelta}
\Omega_2^{(R)}(\ell,\Delta)
=\Omega_2^{(R)}(\ell,1)+z\sum_{i=0}^{\Delta-2}\widetilde{\Omega}_2(\ell,i),
\ee

\begin{figure}
\centering
\includegraphics[width=0.5\columnwidth]{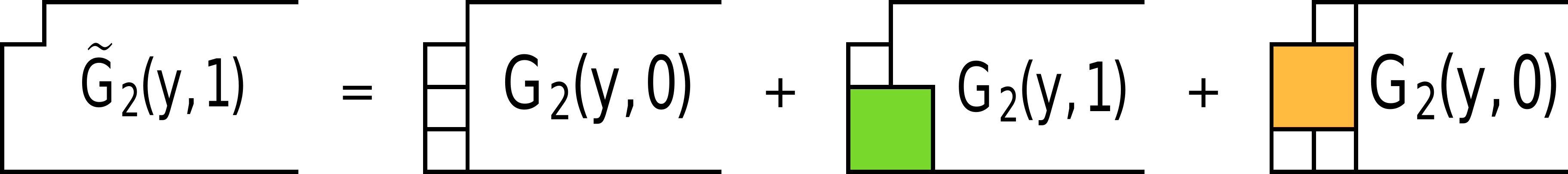}
\caption{Diagrammatic representation of the recursion relation obeyed by the generating function
$\widetilde{G}_2(y,1)$ [see \eref{gty2d} for definition] for a track of width 4. Right hand side enumerates the different ways the first 
column of the track may be occupied by vacancies (open 1$\times$1 square),
square (filled 2$\times$2 square of color green) and defect (filled 2$\times$2 square of color yellow).}
\label{fig:2}      
\end{figure}

To solve \eref{omega2Relldelta}, consider the generating function $\widetilde{G}_2(y,\Delta)$  
defined as
\be
\label{gty2d}
\widetilde{G}_2(y,\Delta)=\sum_{\ell=0}^\infty \widetilde{\Omega}_2(\ell,\Delta)y^{2\ell+3\Delta/2},
\ee
where power of $\sqrt{y}$ gives total number of sites in the system.
The diagrammatic  representation of the recursion relation obeyed by 
$\widetilde{G}_2(y,1)$ is shown in \fref{fig:2} and may be written as
\be
\label{gty21}
\widetilde{G}_2(y,1)=y^{3/2}{G}_2(y,0)+zy^{5/2} G_2(y,1)+z_Dy^{7/2} G_2(y,0),
\ee
where $z_D$ is the activity associated with each defect, and $G_2(y,0)$ and $G_2(y,1)$ are as in 
\eref{sg20} and \eref{sg21}. 
The generating function $\widetilde{G}_2(y,1)$ is then easily solved  
to give
\be
\label{sgty21}
\widetilde{G}_2(y,1)=\frac{(1+z_Dy^2-zz_Dy^4)y^{3/2}}{f(y^2)}.
\ee
For large $\ell$ the partition function may be written asymptotically as
\be
\label{app_at2d}
\widetilde{\Omega}_2(\ell,\Delta)=\widetilde{a}_2(\Delta)\lambda_2^{2\ell+\Delta},~\Delta \geq 0,~ \ell \gg 1.
\ee
Calculating the coefficient of $y^{2\ell+3/2}$ from \eref{sgty21} and using \eref{app_at2d}, we obtain 
the prefactor
\be
\label{at1}
\widetilde{a}_2(1)=\frac{-(1+z_Dy_2-zz_Dy_2^2)}{\sqrt{y_2}f^\prime(y_2)}.
\ee

\begin{figure}
\centering
\includegraphics[width=0.5\columnwidth]{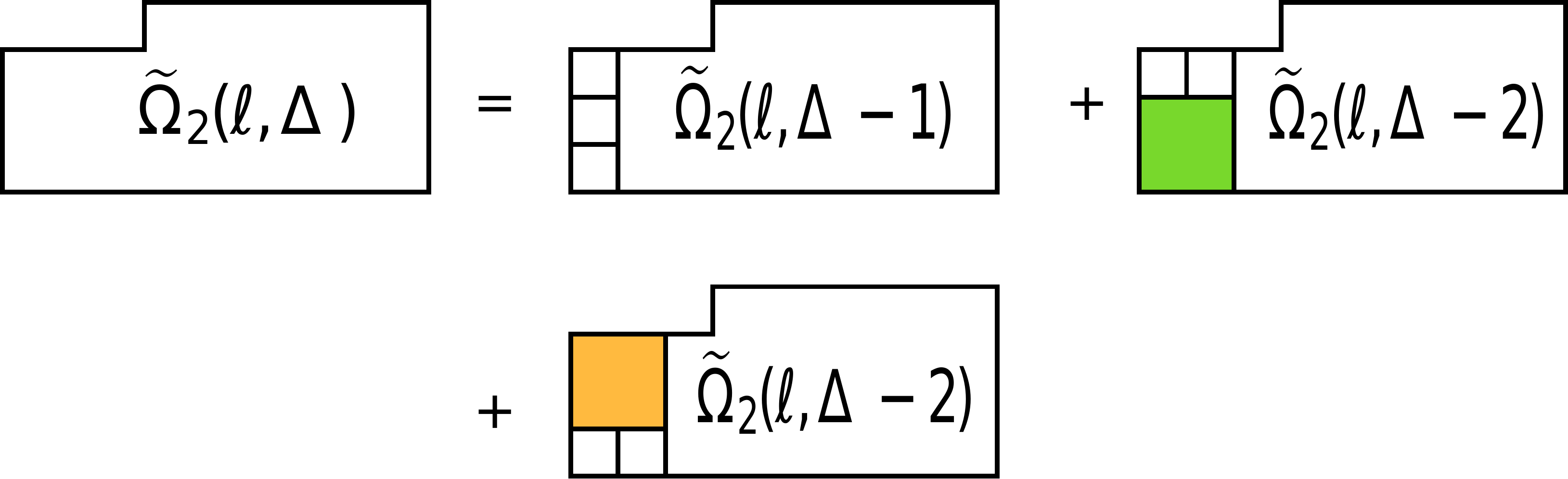}
\caption{Diagrammatic representation of the recursion relation obeyed by the partition function
$\widetilde{\Omega}_2(\ell,\Delta)$ with $\Delta\geq2$ for a track of width 4. Right hand side enumerates 
the different ways the first column of the track may be occupied by vacancies (open 1$\times$1 square),
square (filled 2$\times$2 square of color green) and defect (filled 2$\times$2 square of color yellow).}
\label{fig:5}      
\end{figure}

Now calculate the partition function $\widetilde{\Omega}_2(\ell,\Delta)$ for $\Delta\geq2$. The diagrammatic representation of 
the recursion relation obeyed by the partition function $\widetilde{\Omega}_2(\ell,\Delta)$ 
for $\Delta\geq2$ is shown in \fref{fig:5} and may be written mathematically as
\be
\label{otld}
\widetilde{\Omega}_2(\ell,\Delta)=\widetilde{\Omega}_2(\ell,\Delta-1)+(z+z_D)\widetilde{\Omega}_2(\ell,\Delta-2),
~\Delta=2, 3, ....
\ee
We define the generating function
\be
\label{hlt}
H(\ell,t)=\sum_{\Delta=0}^\infty \widetilde{\Omega}_2(\ell,\Delta)t^{\Delta}.
\ee
Multiplying \eref{otld} by $t^\Delta$ and performing summation over $\Delta$ from $2$ to $\infty$, we obtain a linear equation  
obeyed by $H(\ell,t)$ which is solved to give
\be
H(\ell,t)=\frac{\widetilde{\Omega}_2(\ell,0)+t\left[\widetilde{\Omega}_2(\ell,1)-\widetilde{\Omega}_2(\ell,0)\right]}{1-t-(z+z_D)t^2}.
\ee
$H(\ell,t)$ has two simple poles determined by the roots of the quadratic
equation $1-t-(z+z_D)t^2=0$
\be
t_\pm=\frac{-1\pm\sqrt{1+4(z+z_D)}}{2(z+z_D)}.
\ee
Expanding the denominator about $t_\pm$ and calculating the coefficient of $t^{\Delta}$,
we get the expression for $\widetilde{\Omega}_2(\ell,\Delta)$ and using \eref{app_at2d} the 
prefactor is obtained to be
\be
\label{ta2delta}
\widetilde{a}_2(\Delta)=B_+(t_+\lambda_2)^{-\Delta}+B_-(t_-\lambda_2)^{-\Delta}, ~\Delta\geq 0,
\ee
where
\be
B_\pm=\frac{\pm\bigg[\lambda_2 \widetilde{a}_2(1)-[(z+z_D)t_\mp +1]a_2(0)\bigg]}{\sqrt{1+4(z+z_D)}}.
\ee

We now return to  \eref{omega2Relldelta} and replace the partition functions $\Omega_2^{(R)}(\ell,\Delta)$ and
$\widetilde{\Omega}_2(\ell,i)$ by their  asymptotic forms given in   \eref{lrpartitions2} and \eref{app_at2d} respectively, 
and do the summation 
over $\widetilde{\Omega}_2(\ell,i)$ from $i=0$ to $(\Delta-2)$, to obtain the prefactor
\be
a_2^{(R)}(\Delta)=v_1\lambda_2^{-\Delta}+v_2(t_+\lambda_2)^{-\Delta}+v_3(t_-\lambda_2)^{-\Delta},~\Delta\geq2,
\label{eq:92}
\ee
where
\bea
v_1&=&a_2^{(R)}(1)\lambda_2+z\bigg(\frac{B_+t_+}{t_+-1}+\frac{B_-t_-}{t_--1}\bigg),\nonumber\\
v_2&=&-\frac{zB_+t_+^2}{t_+-1},\nonumber\\
v_3&=&-\frac{zB_-t_-^2}{t_--1}.\nonumber
\eea

\section{\label{sec:6}Results}

In this section we determine the interfacial tension $\sigma(z)$ between two ordered phases
as a function of the activity $z$. From \eref{sigma1},  $\sigma(z)$ may be written as
\be
\label{sigma}
\sigma(z)=-\frac{1}{2}\log\bigg[\frac{\lambda_1^2\Lambda_2}{z a_1 a_2(0)}\bigg],
\ee
where $\Lambda_2$, $\lambda_1$,  $a_1$ and $a_2(0)$ are as in \eref{largest_lambda_s}, 
\eref{lmda1a1},  and \eref{aa20}.
$\Lambda_2$ depends on $a_2^{(L)}(\Delta)$ and $a_2^{(R)}(\Delta)$, which in turn have 
been calculated in \eref{a2_left} and \eref{eq:92}. 
We also set $z_D=z$, where $z_D$ is the activity of a defect.

\begin{figure}
\centering
\includegraphics[width=0.5\columnwidth]{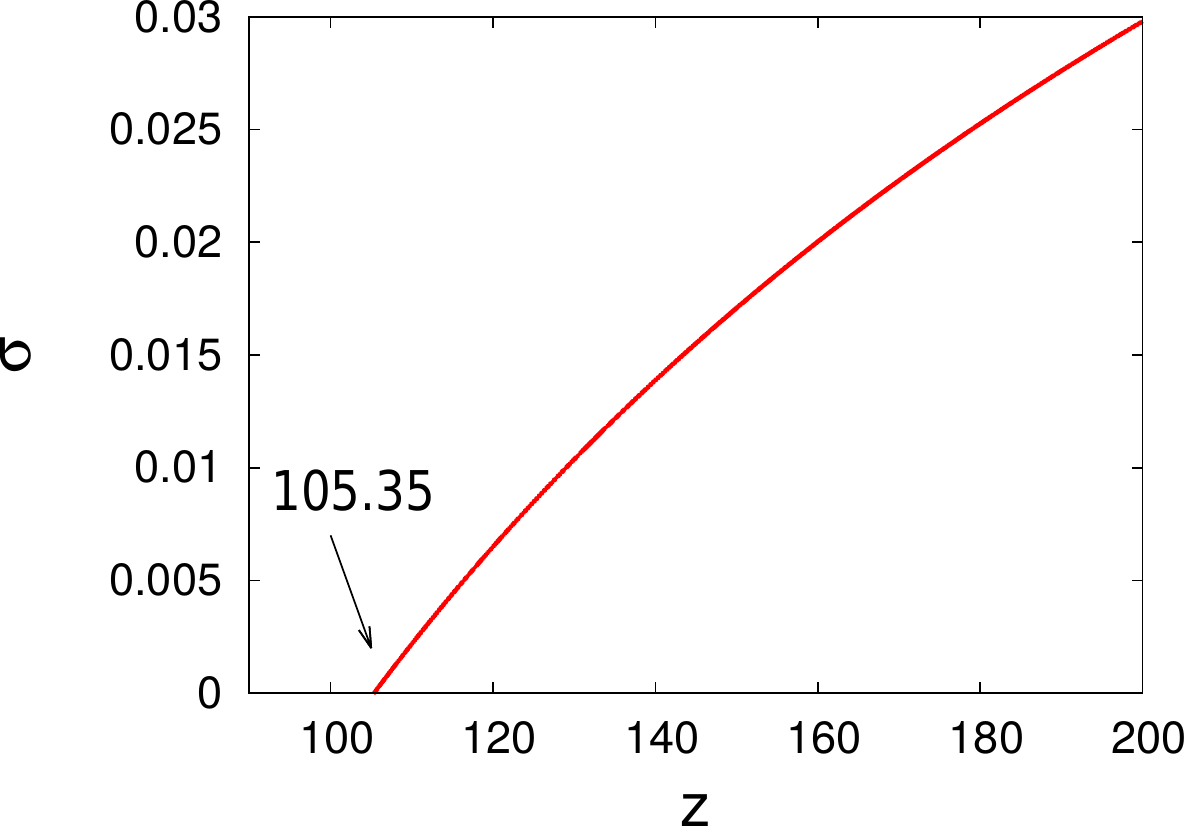}
\caption{The variation of the interfacial tension $\sigma(z)$ with activity $z$. 
Interfacial tension $\sigma(z)$ vanishes at the critical activity $z=z_c$.}
\label{fig:10}      
\end{figure}

The variation of $\sigma(z)$ with activity $z$ is shown in \fref{fig:10}. It decreases monotonically with decreasing $z$
and becomes zero at a finite value of $z$, which will be our estimate of the critical activity $z_c$. We find
that $z_c=105.35$ for the interface with overhangs. 
As a check for the calculation, we confirm that if we set  $z_D=0$, then we obtain the results for the 
estimated $z_c$ in the absence of defects~\cite{trisha2}.  The result for $z_c$ compares well with the numerical
estimate from Monte Carlo simulations of $z_c\approx 97.5$ [see \tref{table:estimates}].

The occupied area fraction or density $\rho$ may be calculated from the partition function $Z^{(0)}$ as:
\be
\label{density}
\rho=\frac{4 z}{N_xN_y}\frac{\partial}{\partial z}\bigg[\log \big(Z^{(0)}\big)\bigg],
\ee
where the factor $4$ accounts for the area of a square.
Substituting for $Z^{(0)}$ from \eref{z0}, the density $\rho$ in  \eref{density}, in the thermodynamic
limit $N_x\rightarrow\infty,~N_y\rightarrow\infty$, reduces to
\be
\label{density_exp}
\rho=4z\bigg[\frac{1}{\lambda_2}\frac{\partial\lambda_2}{\partial z}-\frac{1}{2\lambda_1}\frac{\partial\lambda_1}{\partial z}\bigg].
\ee
We thus obtain the critical density to be $\rho_c=0.947$. This
estimate compare well with the Monte Carlo results of $\rho_c\approx 0.932$ [see \tref{table:estimates}].

\section{\label{sec:7}Conclusion}

In this paper, we estimated the transition point of the disordered-columnar transition in  
in the hard square model by calculating the interfacial tension between two ordered phases within a pairwise approximation.
This calculation allows for multiple defects to be present as well as the interface to have effective overhangs.
We obtain the critical activity $z_c=105.35$ and critical density $\rho_c=0.947$, which agrees reasonably with 
the numerically obtained
results of $z_c\approx 97.5$ and $\rho_c\approx 0.932$. Our estimate for the critical activity is a considerable
improvement over earlier estimates based on many different  approaches [see \tref{table:estimates}].

We calculated the  prefactor $a_2^{(R)}(\Delta)$ by allowing defects to be present as overhangs [see \sref{sec:secar}].
The calculation can be repeated when defects are present only in regions which do not correspond to overhangs.
This corresponds to a defect in the right phase being present only to the  right of $\max(\xi_i,\xi_{i+1})$ 
[see \fref{fig:7}].  This calculation leads to an estimate of $z_c=43.28$, which is about half the value of 
the numerical result of $z_c\approx 97.5$. 
The decrease in the value of $z_c$ on excluding overhangs is consistent with the fact that
the entropy of the system with interface decreases while the entropy of the system without interface remains 
unchanged. We, thus,  conclude that the presence of overhangs in the interface is important
for the calculation of interfacial tension.

A similar analysis for determining the phase boundary  may be done for other kind of systems,
which show a transition from disordered to columnar ordered phase with increasing
density. The mixture of hard squares and dimers~\cite{kabir2} shows such a transition, and so does
the system of $(d\times2)$ hard rectangles~\cite{joyjit3,trisha4,trisha2}. It would be interesting to see whether
the approximation scheme used in this paper is useful in obtaining  reliable 
estimates for the phase boundaries in these problems.

\section*{Acknowledgments}
We thank Deepak Dhar for helpful discussions.

\section*{References}
\bibliographystyle{iopart-num}

\begin{thebibliography}{10}
\expandafter\ifx\csname url\endcsname\relax
  \def\url#1{{\tt #1}}\fi
\expandafter\ifx\csname urlprefix\endcsname\relax\def\urlprefix{URL }\fi
\providecommand{\eprint}[2][]{\url{#2}}

\bibitem{domb}
Domb C 1958 {\em Nuovo Cimento\/} {\bf 9} 9--26

\bibitem{bellerman}
Bellemans A and Nigam R~K 1967 {\em J. Chem. Phys.\/} {\bf 46} 2922--2935

\bibitem{hoover}
Hoover W~G and Rocco A~G~D 1962 {\em J. Chem. Phys.\/} {\bf 36} 3141--3162

\bibitem{kinzel}
Kinzel W and Schick M 1981 {\em Phys. Rev. B\/} {\bf 24}(1) 324--328

\bibitem{amarkaski}
Amar J, Kaski K and Gunton J~D 1984 {\em Phys. Rev. B\/} {\bf 29} 1462--1464

\bibitem{francis2}
Ree F~H and Chesnut D~A 1967 {\em Phys. Rev. Lett.\/} {\bf 18}(1) 5--8

\bibitem{nisbet}
Nisbet R and Farquhar I 1974 {\em Physica\/} {\bf 76} 283 -- 294

\bibitem{fernandez}
Fernandes H~C~M, Arenzon J~J and Levin Y 2007 {\em J. Chem. Phys.\/} {\bf 126}
  114508

\bibitem{kabir2}
Ramola K, Damle K and Dhar D 2015 {\em Phys. Rev. Lett.\/} {\bf 114}(19) 190601

\bibitem{feng}
Feng X, Bl\"ote H~W~J and Nienhuis B 2011 {\em Phys. Rev. E\/} {\bf 83}(6)
  061153

\bibitem{zhitomirsky}
Zhitomirsky M~E and Tsunetsugu H 2007 {\em Phys. Rev. B\/} {\bf 75}(22) 224416

\bibitem{baxter1}
Baxter R~J 1980 {\em J. Phys. A\/} {\bf 13} L61

\bibitem{bellerman2}
Bellemans A and Nigam R~K 1966 {\em Phys. Rev. Lett.\/} {\bf 16}(23) 1038--1039

\bibitem{kabir1}
Ramola K and Dhar D 2012 {\em Phys. Rev. E\/} {\bf 86}(3) 031135

\bibitem{lafuente}
Lafuente L and Cuesta J~A 2003 {\em J. Chem. Phys.\/} {\bf 119} 10832--10843

\bibitem{lafuente2}
Lafuente L and Cuesta J~A 2002 {\em J. Phys. Condens. Matter\/} {\bf 14} 12079

\bibitem{slotte}
Slotte P~A 1983 {\em J. Phys. C\/} {\bf 16} 2935

\bibitem{trisha2}
Nath T, Dhar D and Rajesh R 2016 {\em Europhys. Lett.\/} {\bf 114} 10003

\bibitem{heitor}
Marques~Fernandes H~C, Levin Y and Arenzon J~J 2007 {\em Phys. Rev. E\/} {\bf
  75}(5) 052101

\bibitem{temperley2}
Temperley H~N~V 1961 {\em Proc. Phys. Soc.\/} {\bf 77} 630

\bibitem{degennes}
de~Gennes P and Prost J 1995 {\em The physics of liquid crystals\/} ({\em
  International series of monographs on physics\/} vol~23) (Oxford University
  Press)

\bibitem{adsorbed-on-ni}
Bak P, Kleban P, Unertl W~N, Ochab J, Akinci G, Bartelt N~C and Einstein T~L
  1985 {\em Phys. Rev. Lett.\/} {\bf 54}(14) 1539--1542

\bibitem{chlorine}
Taylor D~E, Williams E~D, Park R~L, Bartelt N~C and Einstein T~L 1985 {\em
  Phys. Rev. B\/} {\bf 32}(7) 4653--4659

\bibitem{bromine}
Mitchell S, Brown G and Rikvold P 2001 {\em Surf. Sci.\/} {\bf 471} 125 -- 142

\bibitem{oxygen}
Zhang Y, Blum V and Reuter K 2007 {\em Phys. Rev. B\/} {\bf 75}(23) 235406

\bibitem{koper}
Koper M~T 1998 {\em J. Electroanal. Chem.\/} {\bf 450} 189 -- 201

\bibitem{joyjit2}
Kundu J and Rajesh R 2014 {\em Phys. Rev. E\/} {\bf 89}(5) 052124

\bibitem{joyjit_rectangle_odd}
Kundu J and Rajesh R 2015 {\em Euro. Phys. J. B\/} {\bf 88} 133

\bibitem{joyjit3}
Kundu J and Rajesh R 2015 {\em Phys. Rev. E\/} {\bf 91}(1) 012105

\bibitem{trisha4}
Nath T, Kundu J and Rajesh R 2015 {\em J. Stat. Phys.\/} {\bf 160} 1173--1197

\bibitem{interacting_dimer}
Alet F, Ikhlef Y, Jacobsen J~L, Misguich G and Pasquier V 2006 {\em Phys. Rev.
  E\/} {\bf 74}(4) 041124

\bibitem{trisha1}
Nath T and Rajesh R 2014 {\em Phys. Rev. E\/} {\bf 90}(1) 012120

\bibitem{trisha3}
Nath T and Rajesh R 2016 {\em J. Stat. Mech.\/} {\bf 2016} 073203

\bibitem{quantum_dimer}
Papanikolaou S, Luijten E and Fradkin E 2007 {\em Phys. Rev. B\/} {\bf 76}(13)
  134514

\bibitem{quantum_dimer2}
Ralko A, Poilblanc D and Moessner R 2008 {\em Phys. Rev. Lett.\/} {\bf 100}(3)
  037201

\bibitem{quantum_dimer3}
Wenzel S, Coletta T, Korshunov S~E and Mila F 2012 {\em Phys. Rev. Lett.\/}
  {\bf 109}(18) 187202

\bibitem{s_jin}
Jin S and Sandvik A~W 2013 {\em Phys. Rev. B\/} {\bf 87}(18) 180404

\bibitem{baxter-comb}
Baxter R~J 1999 {\em Ann. Comb.\/} {\bf 3} 191--203

\bibitem{squares_torus}
Blair D~W, Santangelo C and Machta J 2012 {\em J. Stat. Mech.\/} {\bf 2012}
  P01018

\bibitem{packing_square}
Decaudin P and Neyret F 2004 {\em Eurographics\/}  49--52

\bibitem{brownian_square}
Zhao K, Bruinsma R and Mason T~G 2011 {\em Proc. Natl. Acad. Sci.\/} {\bf 108}
  2684--2687

\bibitem{vibrating_square}
Walsh L and Menon N 2016 {\em J. Stat. Mech.\/} {\bf 2016} 083302

\bibitem{kundu}
Kundu J, Rajesh R, Dhar D and Stilck J~F 2012 {\em AIP Conf. Proc.\/} {\bf
  1447} 113--114

\bibitem{joyjit1}
Kundu J, Rajesh R, Dhar D and Stilck J~F 2013 {\em Phys. Rev. E\/} {\bf 87}(3)
  032103

\end{thebibliography}
\providecommand{\newblock}{}

\end{document}